# D-CryptO: Deep learning-based analysis of colon organoid morphology from brightfield images

Lyan Abdul[1], Jocelyn Xu[2], Alexander Sotra[1], Abbas Chaudary[3], Jerry Gao[4], Shravanthi Rajasekar[3], Nicky Anvari[1], Hamidreza Mahyar[5], and Boyang Zhang[1,3*]

[1]School of Biomedical Engineering, McMaster University, 1280 Main Street West, Hamilton, ON, L8S 4L8, Canada
[2]Faculty of Engineering, McMaster University, 1280 Main Street West, Hamilton, ON, L8S 4L8, Canada
[3]Department of Chemical Engineering, McMaster University, 1280 Main Street West, Hamilton, ON, L8S 4L8, Canada
[4]Faculty of Science, McGill University, 845 Sherbrooke Street West, Montreal, QC H3A 0G4, Canada
[5]W Booth School of Engineering Practice and Technology, McMaster University, 1280 Main Street West, Hamilton, ON, L8S 4L8, Canada

*Correspondence to zhangb97@mcmaster.ca

**Abstract**

Stem cell-derived organoids are a promising tool to model native human tissues as they resemble human organs functionally and structurally compared to traditional monolayer cell-based assays. For instance, colon organoids can spontaneously develop crypt-like structures similar to those found in the native colon. While analyzing the structural development of organoids can be a valuable readout, using traditional image analysis tools makes it challenging because of the heterogeneities and the abstract nature of organoid morphologies. To address this limitation, we developed and validated a deep learning-based image analysis tool, named D-CryptO, for the classification of organoid morphology. D-CryptO can automatically assess the crypt formation and opacity of colorectal organoids from brightfield images to determine the extent of organoid structural maturity. To validate this tool, changes in organoid morphology were analyzed during organoid passaging and short-term forskolin stimulation. To further demonstrate the potential of D-CryptO for drug testing, organoid structures were analyzed following treatments with a panel of chemotherapeutic drugs. With D-CryptO, subtle variations in how colon organoids responded to the different chemotherapeutic drugs were detected, which suggest potentially distinct mechanisms of action. This tool could be expanded to other organoid types, like intestinal organoids, to facilitate 3D tissue morphological analysis.

**Introduction**

Monolayer cell-based assays are an invaluable tool for studying cellular functions *in vitro*. However, these models do not accurately recapitulate *in vivo* tissue responses. This is largely because monolayer cell models do not exhibit tissue-specific architecture and lack the appropriate 3D cellular microenvironment. Stem cell-derived organoids that can spontaneously differentiate and self-assemble into 3D tissues with structures that resemble many features of the native organ have emerged as alternative *in vitro* models.[1] For instance, colon organoids have been widely used as large intestine models due to their structural and functional similarities.[2] An important feature of the colon epithelium is the crypt, which are epithelial invaginations that renew the intestinal lining every 3-5 days.[3] The organization of the crypt is crucial for the regeneration of the epithelium *in vivo*. Stem cells at the base of the crypt are protected from continuous mechanical and chemical stressors, and as a result, can proliferate and differentiate to regenerate the epithelium. Similarly, colon organoid morphology reflects the structure and organization of the native colon crypts by exhibiting budding structures which contain the stem cells that give rise to colon-specific cells.[4,5] Therefore, analyzing organoid morphology can provide insights into colon physiology and pathophysiology *in vivo*.

Qualitative analysis of colon organoid morphology, specifically the opacity and budding of organoids, has largely been used to assess the maturity of colon organoids. Colon organoids that are more transparent, have thinner walls, and are cystic are indicative of an earlier differentiation state.[6] On the other hand, colon organoids have reached a more differentiated state when they are more opaque due to the thickening of the epithelial wall. Differentiated colon organoids also exhibit more budding structures that resemble the colon crypt which is the stem cell niche that controls colonocyte renewal and homeostasis.[2] Previously, the presence of budding within small intestinal organoids has been used to optimize the extracellular matrix, study stem cell differentiation, and understand the mechanics of epithelial folding. [3,7–9] Analysis of budding has also been used to study diseases. For example, colon organoids from individuals with inflammatory bowel disease or tumour-derived organoids had lower rates of budding structures.[10,11] However, to assess these morphological differences, previous work used manual analysis or relied on traditional image analysis that uses imperfect parameters such as eccentricity to describe organoid shapes.[12–14]

To facilitate the morphological analysis of organoids with abstract features that are not easily defined by traditional image analysis parameters, a type of computer vision called deep learning can be applied. Deep learning refers to an automated method of computer-based image recognition that relies on using pre-existing data to make predictions on new image instances.[15] Traditional computer recognition techniques rely on manual feature extraction to distinguish between the categories of interest. With deep neural networks, both feature extraction and classification are done automatically without any input from the user. This provides several advantages. First, colon organoid features are learned directly from the images without the need for manual feature extraction. Second, analysis of the structures is not limited to using shape descriptors, so organoid morphology can be characterized despite the high heterogeneity of colon organoid structure. Third, automatic image analysis can improve the throughput of morphological analysis. Finally, these models could be trained to correctly classify between categories despite imaging artifacts. Artificial neural networks have been previously used to detect and count intestinal organoids and replace immunostaining and cell viability assays. [16–19] However, deep learning has yet to be used for the morphological characterization of any type of organoids.

Hence, we used deep learning to characterize the morphological structure of organoids by developing an analysis tool, D-CryptO, to distinguish between transparent and opaque organoids, as well as spherical and budding organoids. Collectively, these features reveal the structural maturity and health of colon organoids. To validate our deep learning model, we analyzed changes in colon organoid morphologies in (1) organoid passaging, (2) short-term forskolin stimulation, (3) a drug screening study with a panel of six chemotherapeutic drugs, and (4) a dose-response study to doxorubicin. We found that morphological analysis allowed us to capture variations in how colon organoids responded to the different chemotherapeutic drugs, which provide insights into the potential mechanisms of drug toxicity.

**Results**

**Colon organoid culture and morphological characteristics**

Colorectal organoids, derived from primary colon tissue, were embedded in Matrigel, and cultured for a period of 7 days in a 24-well plate **(Figure 1a-b)**. The primary tissue contains adult stem cells which proliferate and differentiate to form the colon organoids *in vitro*. To determine organoid maturation, we performed histological analysis of the colon organoids. Organoids expressed villin apically, a marker for microvilli, which is indicative of differentiated intestinal cells. Furthermore, the expression of ki-67 indicated that stem cells were also present within the colon organoids (**Figure 1c**). We observed a spectrum of morphologies from these colon organoids. Organoids differed in their opacity as well as the extent of budding **(Figure 1d)**. Colonospheres are transparent with little-to-no budding. On the other hand, colonoids are more mature organoids that are opaque with a significant number of budding structures.[20,21] As the proliferating stem cells differentiate into organ-specific cells, opacity increases due to changes in epithelium thickness. [7] We also observed organoids that exhibited some characteristics of both colonospheres and colonoids. For example, some organoids were spherical and opaque while other organoids had buds and were transparent. Hence, using the parameters of both opacity and budding could give an indication of the structural maturity of the organoids grown *in vitro.*

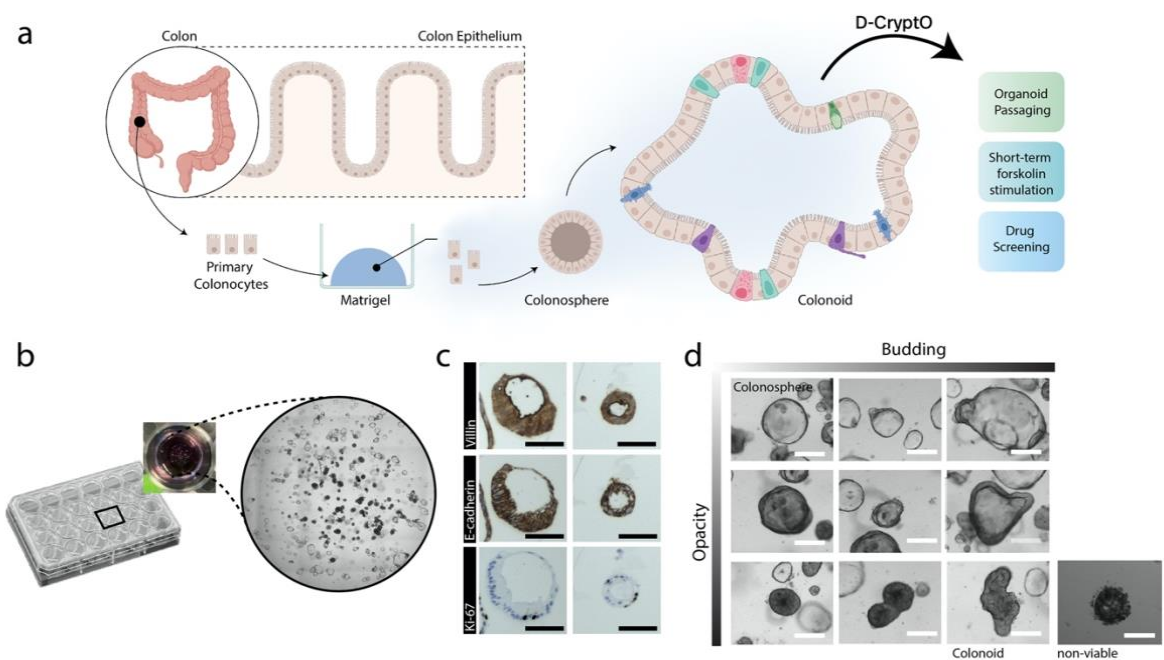

**Figure 1. Morphological heterogeneity of colon organoids. a,** Illustration of primary cells embedded in Matrigel® that self-assemble into colonospheres and develop into colonoids. **b,** Organoids embedded in 50µL of Matrigel® in a standard 24-well plate. **c,** Histological sections of organoids expressing the mature markers of villin, E-cadherin and ki-67. Scale bar, 100 µm. **d,** Representative images of organoids exhibiting varying levels of opacity and budding. Scale bar, 200 µm

**Dataset creation and model training**

To analyze these parameters using deep learning, we first created two custom datasets using images of individual organoids. Organoid images were obtained by taking montages and z-stack images of the entire well (**Figure 2a**). Using these images, organoids were sorted into the first dataset, which consisted of examples of budding and spherical organoids. The second dataset contained examples of opaque and transparent organoids **(Figure 2b)**. The dataset for opacity contained 1021 images of opaque organoids and 1457 images of transparent organoids. The dataset for the budding feature contained 1081 images of budding organoids and 1395 images of non-budding organoids. These datasets were further split into training, validation, and test datasets. We made the full training dataset publicly accessible at the open science framework data repository: https://osf.io/42r3g/. Next, we fine-tuned six pre-trained deep neural network models (ResNet152V2, XCeption, InceptionResNetV2, VGG-16, VGG-19, ResNet50) for each parameter using the custom datasets.[22–25] These models were selected either because they had higher speeds or performed more accurately on the ImageNet dataset. We implemented these transfer learning approaches using the Keras framework with the Tensorflow backend.[26] For opacity, both XCeption and VGG-16 performed with an accuracy of 98% on the test set. For budding, both ResNet152V2 and XCeption performed with an accuracy of 90.87% on the test set. XCeption, a convolutional neural network model, was chosen as the final model since it performed most accurately for both parameters **(Figure 2c)**. To understand which regions of the organoid were used for classification, heat maps were outputted to highlight important locations **(Figure 2d)**. For the opacity model, the center of the organoid is important for distinguishing between opaque and transparent organoids. For the budding model, the edges of the organoids are used to distinguish between budding and spherical organoids. For opacity, there was a lower rate of false positives and negatives compared to budding **(Figure 2e, f)**. This could be due to the lower heterogeneity in the opacity of organoids compared to the budding morphologies. Additionally, overlapping organoids can affect classification accuracy (**Supplementary Figure 2a**). While D-CryptO can correctly classify overlapping organoids, there are cases where organoids are misclassified. This is especially the case when one organoid has several overlapping organoids along its perimeter, which D-CryptO occasionally misclassified as having budding structures. Additionally, if a transparent organoid without buds overlaps an opaque organoid, it is at times misclassified as being a budding organoid. On the other hand, opaque and budding organoids are not affected if other organoids overlap with them (**Supplementary Figure 2b**). Overall, overlapping organoids impact the classification of opacity very little, as D-CryptO performed with an accuracy of 96%, while the classification of the budding feature was impacted with D-CryptO obtaining a classification accuracy of 68% (**Supplementary Figure 2c**). However, in both 384-well and 24-well plates, the percentage of overlapping organoids is 16% and 20% respectively, which is quite low (**Supplementary Figure 2d**). Nonetheless, the overall accuracy was above 85% for both parameters and both models had low rates of false negatives. Together, these trained models were combined to become D-CryptO for analyzing colon organoid morphology and determining the extent of colon organoid maturation. D-CryptO can morphologically analyze a large volume of organoids within a short time period. For 1 organoid, D-CryptO took 1 second to output its classification scores for both categories. When the organoid number was increased to 1000, it took 17 seconds to output its classification scores. Hence, D-CryptO's speed makes it suitable for high-throughput drug analysis (**Figure 2g**). Lastly, we showed that D-CryptO can successfully capture the extent of differences in morphology in both budding and opacity characteristics, with a prediction score that reflects where the organoid falls on the spectrum of budding or opacity (**Figure 2h, i**).

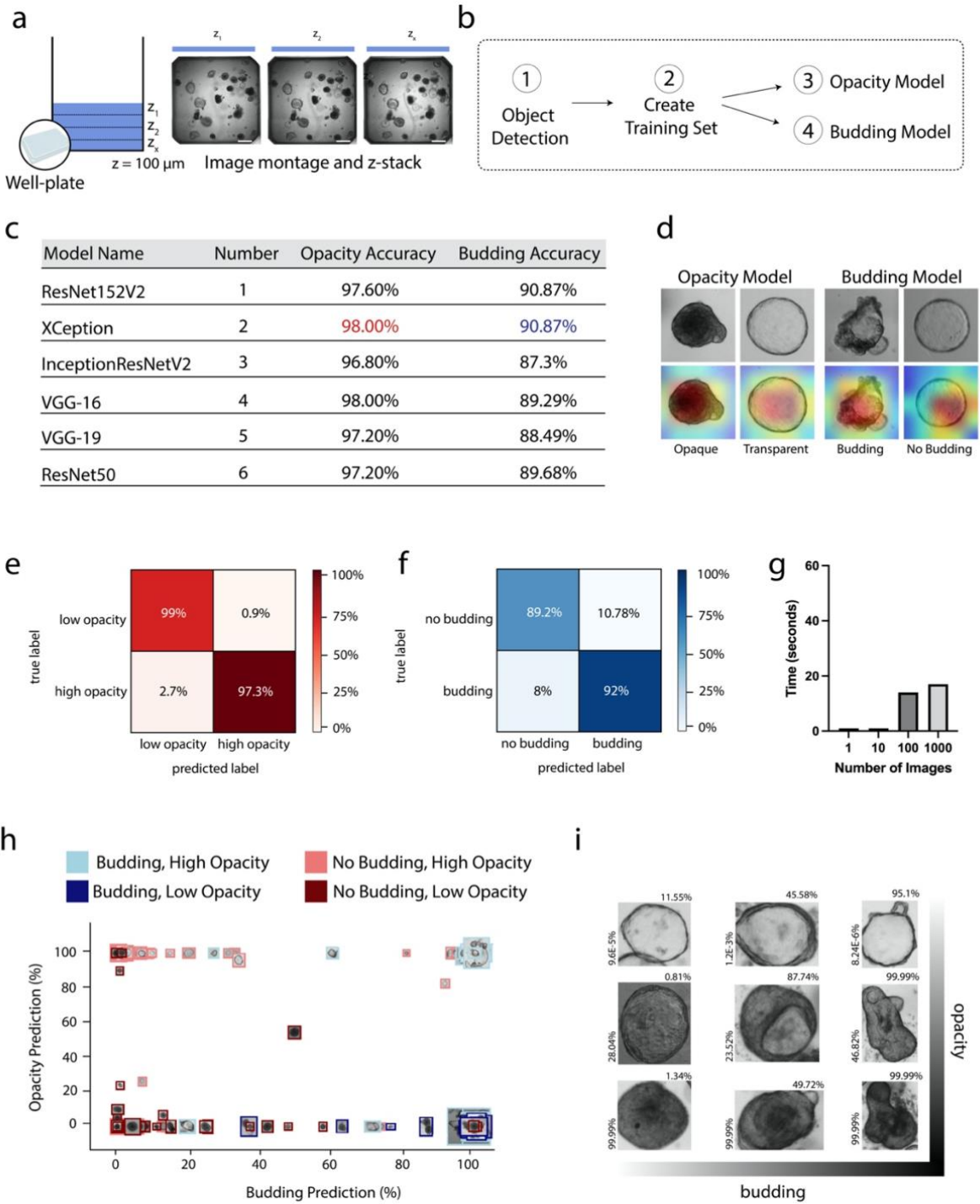

**Figure 2. D-CryptO training and testing. a,** Image acquisition workflow. **b,** Image analysis workflow. **c,** Accuracy of trained models on test set following transfer learning. **d,** Heatmaps identifying which parts of the image were more important for organoid classification (red indicates higher importance and blue indicates lower importance). **e,** Confusion matrix for the opacity feature of D-CryptO. **f,** Confusion matrix for the budding feature of D-CryptO. **g,** Time required for organoid morphological analysis. **h,** Organoid distribution based on D-CryptO predictions for opacity and budding. **i,** Representative D-CryptO organoid classification prediction score and corresponding brightfield images.

**Morphological changes of organoids during organoid expansion and passaging**

To validate D-CryptO, we used it to analyze organoid morphology in several different case studies. While colon organoids are a valuable tool for biological applications, there is a lot of variability in morphology within a single Matrigel dome. This can hinder the reproducibility of experimental results during culture. We analyzed opacity and budding to determine variability in morphology over prolonged culture **(Figure 3a,b)**. First, we analyzed organoids after thawing them directly into a 24-well plate and culturing them for 5 days. Next, an image montage and three z-stack sections were obtained from the Matrigel domes in the 24-well plate. Then, each individual organoid from these images was automatically cropped and inputted into D-CryptO for analysis. The average percentage of opaque organoids was 21.4 ± 2.9% while the average percentage of budding organoids was 74.5% ± 2.1%. The average organoid diameter was 271.4 μm ± 14.4 μm. Next, we passaged the organoids and repeated the analysis. While there was greater variability in different wells, there wasn't a significant difference in opacity, budding, diameter, or the number of detected organoids following passaging **(Figure 3c,d,e, f)**. This demonstrates that organoids remain robust following one passage, but further analysis is required to see how a greater number of passages impacts organoid morphology. Nonetheless, D-CryptO could be used to analyze colon organoid culture to assess organoid morphology over time non-invasively.

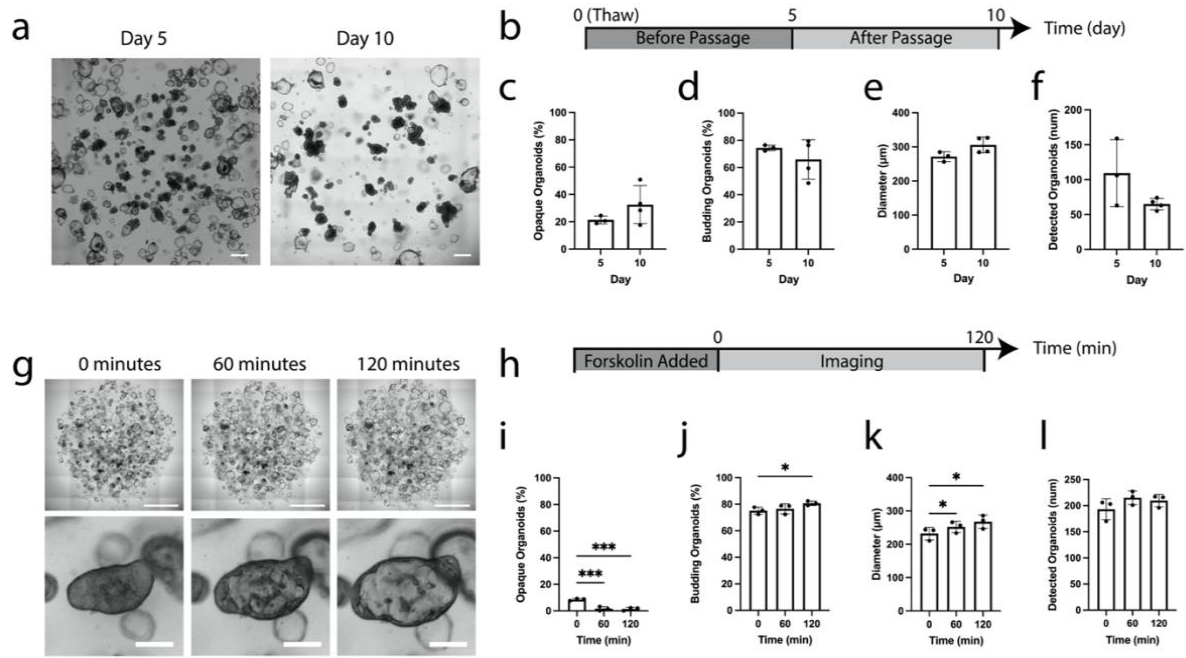

**Figure 3. Organoid morphological changes during passaging and short-term forskolin stimulation. a,** Brightfield images of organoids cultured over two weeks. Scale bar, 500 μm. **b,** Timeline of organoid culture. Organoids were thawed and cultured for 1 week and were subsequently passaged. **c-f,** Quantification of the percentage of opaque organoids, percentage of budding organoids, diameter and the number of organoids (n=3-4) on days 5 and 10. **g,** Brightfield images of organoids during forskolin stimulation. Scale bar (top), 2000 μm. Scale bar (bottom), 100 μm. **h,** Timeline of forskolin treatment. **i-l,** Quantification of the percentage of opaque organoids, budding organoids, diameter, and the number of organoids following 60 and 120 minutes of forskolin treatment (n=3). *$p < 0.05$, **$p < 0.01$, ***$p < 0.001$.

**Morphological changes of organoids to short-term exposure to external stimuli**

Next, we used D-CryptO to assess changes in organoid morphology during short-term perturbation. Colon organoids were thawed and embedded in Matrigel in a 24-well plate and cultured for a period of 10 days. We applied 10 μM of forskolin, a small molecule that activates the cystic fibrosis transmembrane conductance regulator (CFTR), for a period of 2 hours **(Figure 3g,h)**. The CFTR channel is essential for ion transport and mucus production in the colon. In healthy organoids that have a functional CFTR channel, forskolin treatment results in the opening of the channel, the movement of chloride ions through the CFTR channel, and the subsequent flux of water into the organoid. As a result, the organoid swells.

However, organoids with mutations in the CFTR channel do not exhibit this response.[27] To analyze the morphological changes, we first acquired whole-well montages and several z-stacks of each of the wells and then automatically cropped each individual organoid from these images. Using D-CryptO, we analyzed changes in opacity and budding for each organoid in response to forskolin. The percentage of opaque organoids significantly decreased following 60 and 120 minutes of forskolin treatment since the water was accumulating within the organoid lumen (**Figure 3i**). Budding also increased slightly after 120 minutes (**Figure 3j**). This could be due to the budding domains of the organoids becoming more apparent following forskolin stimulation. Diameter also increased following forskolin stimulation, as expected due to organoid swelling **(Figure 3k)**. Similarly, the number of detected organoids remained the same during the 120-minute treatment (**Figure 3l**).

**Morphological changes of organoids in response to drug treatments**

Chemotherapeutic drugs have been shown to induce gastrointestinal toxicity *in vivo* which can affect treatment outcomes.[28] We used D-CryptO to assess the effect of different clinically approved chemotherapeutics at a single dosage on colon organoid morphology. We thawed colon organoids directly into a 384-well plate and applied 6 chemotherapeutic drugs at a concentration of 50 µM to the organoids following 4 and 10 days of culture **(Figure 4a)**. This concentration is higher than the maximum recommended plasma clinical concentrations or has been previously shown to have toxic effects on colon organoids.[29,30] Then, we acquired images of each well in the 384-well plate by taking a montage with z-stacks (14-15 sections separated by 100µm distance) to capture all the organoids within the Matrigel. The entire image was then analyzed using object detection to identify each individual organoid. Each automatically cropped organoid was then inputted into D-CryptO for analysis. We also conducted an LDH (Lactate dehydrogenase) assay to non-invasively assess the cytotoxicity of the chemotherapeutics to supplement the results from D-CryptO. For organoids treated with docetaxel or chlorambucil, the opacity, budding, diameter and number of organoids remained the same over 10 days. Similarly, the LDH absorbance was not significantly different compared to the control. This indicates that these drugs did not have a cytotoxic effect on the organoids (**Figure 4c,d,e,j**). However, it does appear that the morphological development of the organoids was slowed down as the percentage of budding organoids did not increase as in the control condition. In the other drug treatment groups, LDH absorbance was significantly increased compared to control. Similarly, organoid morphology was impacted. For example, in the fluorouracil condition, while the percentage of opaque organoids decreased by day 10, the percentage of budding organoids did not increase. This could indicate that fluorouracil inhibited organoid budding (**Figure 4f**). In the cisplatin-treated organoids, the percentage of budding organoids increased while the percentage of opaque organoids and the organoid diameter remained the same. This could indicate that the cisplatin inhibited organoid growth (**Figure 4g**). Organoids treated with erlotinib showed an increase in the number of opaque organoids, no change in budding, and a decrease in diameter. This could indicate that erlotinib prevented stem cell proliferation and differentiation, which induced organoid collapse (**Figure 4h**). Organoids treated with doxorubicin also resulted in an increase in opaque organoids and budding organoids, while the diameter of the organoids remained the same. When examining the images, it was more apparent that budding did not increase, but that there was a higher percentage of non-viable and dissociating organoids, which were falsely classified as organoids containing budding features. This morphological change could indicate that doxorubicin induced toxicity in colon organoids and triggered cell apoptosis (**Figure 4i**). Therefore, by analyzing all three organoid morphological parameters, budding, opacity and diameter holistically, we can identify cytotoxic agents and gain new insights into the potential mechanisms of drug-induced toxicity.

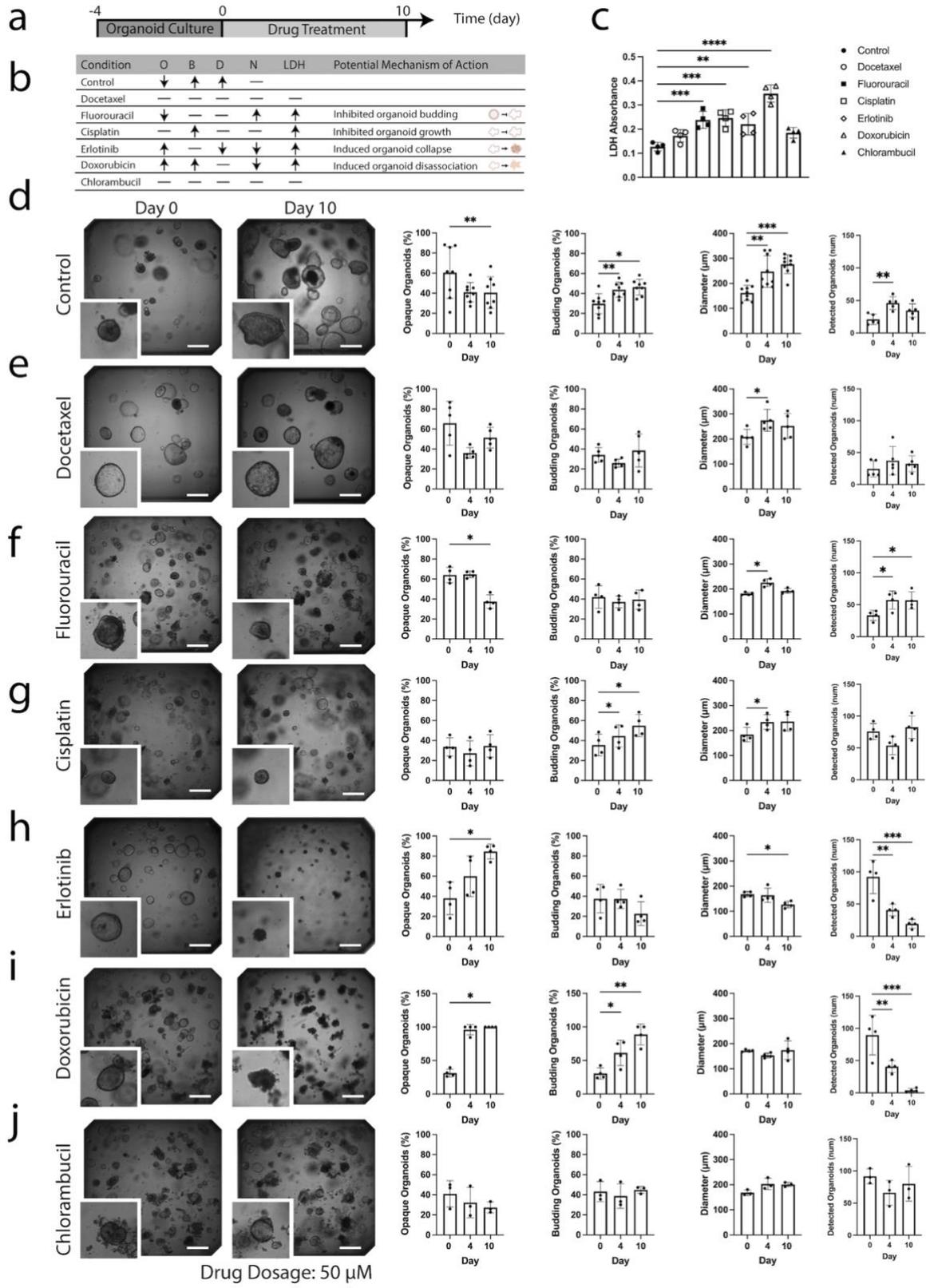

**Figure 4. Chemotherapeutic drug-induced morphological changes in organoids. a,** Timeline of organoid culture and drug treatment. **b,** An outline of drug-induced effects on organoid opacity (O), budding (B), diameter (D), the number of organoids (N), LDH assay results (LDH) and potential mechanisms of action. **c,** LDH absorbance following 4 days of treatment with chemotherapeutic agents. **d,** Brightfield images of organoids in the non-treated

condition and quantification of the changes in opacity, budding, diameter, and the number of organoids following 10 days of culture (n=4). **e-j,** Organoid brightfield images (4× magnification) and quantification of the changes in opacity, budding, diameter, and the number of organoids on days 0, 4, and 10 following treatment with **(e)** docetaxel (n=4), **(f)** fluorouracil (n=4), **(g)** cisplatin (n=4), **(h)** erlotinib (n=4), **(i)** doxorubicin (n=4), and **(j)** chlorambucil (n=3). Scale bar, 500 µm. *$p < 0.05$, **$p < 0.01$, ***$p < 0.001$, ****$p < 0.0001$.

**Discussion**

Analyzing the structural complexity within organoids can provide valuable insights. So far deep learning methods have mainly been used to detect, segment or track organoids.[17,31] We developed a deep-learning based method, D-CryptO, to characterize the complex structural morphology of organoids for the first time. As a result, the extent of organoid maturity can be analyzed automatically without the use of invasive analyses such as immunofluorescent staining. We validated this tool by analyzing changes in organoid morphology in prolonged culture, in short-term perturbation with forskolin, and in chemotherapeutic drug screening to assess drug toxicity. D-CryptO provides several advantages over existing organoid analysis workflows. First, despite the high inter-organoid heterogeneity, D-CryptO accurately categorized organoid opacity and budding. Second, D-CryptO uses brightfield images which allows for non-destructive organoid analysis. Lastly, since D-CryptO makes classifications on single organoids, image analysis can easily be done on organoids grown in a 24-well plate or be scaled up to a 384-well plate as we have shown. D-CryptO can also be further expanded to monitor each organoid's development over time at a single organoid resolution as they transition from colonospheres to colonoids.[14,32] This tracking feature could potentially be used to assign a growth rate for each organoid, and assess how each individual growth rate is impacted following drug treatment as each organoid is composed of a heterogenous cell population. Furthermore, D-CryptO could potentially be adapted to analyze more subtle differences in the opacity and budding of organoids. This could help identify organoids that are more proliferative or are composed of a differentiated cell population. For example, colon organoids can be fluorescently stained to determine the different cell types present and study if this is reflected in their morphology seen in brightfield images. [3,7–9] If there are patterns observed within the brightfield morphologies and the present cell types, D-CryptO could be trained to classify between these morphologies. This potentially could be valuable to gain a more accurate understanding of how drugs differentially impact organoids containing different proportions of stem cells or differentiated cell types.

Despite the advantages of D-CryptO, there are some limitations. First, organoids treated with cytotoxic compounds were identified as having increased budding structures. This is possibly due to cell death aggregations having a similar morphological outline to budding organoids. In our drug studies, as a quality control step we also visually assessed the images and found this classification error. While manual assessment of the images may not be suitable to identify errors in classification following treatment with cytotoxic agents, especially for high-throughput applications, deep learning can also be used to identify dying organoids. For example, organoids can be stained with a viability reagent such as propidium iodide which can then be used to identify viable and non-viable organoids in brightfield images. These brightfield images can then be compiled into a dataset to train a new deep learning model that can distinguish between living and dead organoids. This classifier can be added upstream of D-CryptO so that only viable organoids are passed on for morphological analysis. Furthermore, blurry organoids or overlapping organoids, although part of the training dataset, were occasionally misclassified. The training dataset could be increased to improve classification accuracy. Finally, D-CryptO was specifically trained to assess crypt structures of colon organoids, but it is unclear whether D-CryptO can be applied to organoids from other organoids like the small intestine. While many other types of organoids do also exhibit budding features, more transfer learning might be needed for morphological assessment of other types of organoids.[33]

**Conclusion**

Colon organoid morphology exhibits key features of the colon epithelium *in vivo* and could provide information on colon physiology and pathophysiology. In this work, we developed D-CryptO, a deep learning tool to automatically analyze colon organoid structure. Specifically, D-CryptO can analyze the opacity and the presence of budding within colon organoids to assess the extent of tissue maturation and differentiation. To validate D-CryptO, we used it to analyze colon organoid morphology in several cases. We analyzed changes in organoid morphology during organoid culture, during short-term exposure to forskolin, and in a drug screen with a panel of chemotherapeutic drugs. By using D-CryptO to analyze organoid structure following drug treatment, we gained insights into the potential mechanisms by which the drugs induced toxic effects. D-CryptO can help facilitate the analysis of colon organoid morphology to better understand tissue physiology *in vivo*, assess drug effects, and develop therapies.

**Materials and Methods**

**Colon organoid culture**

Patient-derived colorectal organoids were acquired from the University Health Network (UHN) Princess Margaret Living Biobank in Toronto Canada. Approval for the use of these organoids was obtained from the Hamilton Integrated Research Ethics Board under project number, 5982-T. Organoids were derived from a 69-year-old, female patient. Organoids were cultured by thawing frozen vials and embedding them into growth-factor reduced Matrigel. 50µL of Matrigel and the organoids were cast into a 24-well plate. The organoids were maintained using Intesticult human organoid growth media purchased from Stemcell Technologies (Cat #06010) supplemented with Rock Inhibitor. Organoids were grown for a week and later passaged. To passage organoids, the Matrigel was first degraded by incubating Cell Recovery Solution (Corning, Cat# CACB354253) (1mL per well) for 1 hour. Next, 5 mL of Advanced DMEM/F12 media from Gibco (Cat# 12634010) was added to the solution and centrifuged at 200G for 4 minutes. Following centrifugation, the supernatant was discarded and 1mL of TrypLE express enzyme from Gibco (Cat# 12605010) was added to the organoids. The mixture was incubated in the water bath for 10 minutes. 5 mL of Advanced DMEM/F12 media was then added and the contents were centrifuged at 200G for 4 minutes. The supernatant was removed again, and the organoids were embedded in Matrigel and split into 3 wells. For experiments in a 384-well plate, organoids were thawed directly and embedded in 25µL of Matrigel. Organoids used in these experiments were between passages 18-19.

**Image acquisition**

Brightfield images of colorectal organoids were acquired using a Cytation 5 cell imaging multi-mode reader (BioTek® Instruments). For each experimental well, montages and z-stack sections with a distance of 100 µm between them were taken. Both image montages and z-stacks were captured at 4×magnification. Acquired images were then converted into the png and RGB formats.

**Dataset creation**

A set of image montages composed of 35 images were obtained from organoids cultured in a 24-well plate. Each organoid within the image was labelled using labelImg, and its coordinates were used to automatically crop each organoid. Organoids were then sorted into separate datasets. For the opacity dataset, if the organoid had a thin epithelium or a clear lumen it was classified as transparent. If the organoid had a thicker epithelium or did not have a clear lumen, it was classified as opaque. The opacity training dataset consisted of 816 opaque organoids and 1165 transparent organoids. The opacity validation dataset consisted of 101 opaque organoids and 144 transparent organoids. The opacity test dataset contained 104 opaque organoids and 148 transparent organoids. Images were randomly split into the datasets with a ratio of 80:10:10. For the budding dataset, if an organoid had a clear protrusion it was classified as a budding organoid. If an organoid was mainly spherical, it was classified as non-budding. The budding training dataset contained 979 images of budding organoids and 1245 images of non-budding organoids. The budding test set had 102 images of budding organoids and 150 images of non-budding organoids. The budding validation set was created automatically using Keras with a validation split of 20%. The raw data set can be found at https://osf.io/42r3g/.

**Model architecture and selection**

Six pre-trained models were selected for transfer learning: ResNet152V2, XCeption, InceptionResNetV2, VGG-16, VGG-19, and ResNet50. These models were selected based on their performance on the ImageNet dataset as well as their speed. Additionally, these models have different architectures. VGG-16 contains 16 layers, consisting of convolutional layers and max-pooling layers, followed by a densely connected classifier. VGG19 has a similar architecture but consists of 19 layers. ResNet50 contains 50 layers and uses residual connections to reduce the problem of vanishing gradients and improve accuracy. ResNet152V2 also incorporates residual connections but is a deeper model with 152 layers. XCeption uses depthwise separable convolutions to generate a model with fewer parameters and increase performance. InceptionResNetV2 contains 164 layers and combines the Inception architecture, which includes different convolutional filter sizes and incorporates the residual connections of the ResNet architecture.

**Model configuration and training**

Keras (version 2.8.0) and python 3.7 was used to configure and train all models. First, all images were preprocessed into a tf.data.Dataset. Image size for all models was set to (150 pixels, 150 pixels), the batch size was set to 32 and image pixels were normalized to values between 1 and -1. For the opacity feature of D-CryptO, both feature extraction and fine-tuning were conducted. To do this, each of the 6 models with different architectures (ResNet152V2, XCeption, InceptionResNetV2, VGG-16, VGG-19, and ResNet50) were first instantiated and the pre-trained weights were loaded into them. All layers in the pre-trained models were frozen and a new classifier was added which included a dropout layer (dropout rate of 0.2) and a dense layer with 2 nodes and the softmax activation function. The model was then trained for 20 epochs using the Adam optimizer, the categorical cross-entropy loss function, and the categorical accuracy metric to assess model performance. To further improve model performance for opacity, the models were fine-tuned by unfreezing all layers and the model was retrained at a low learning rate of $1 \times 10^{-5}$ for 10 epochs with the same loss function and accuracy metric used for feature extraction. For the budding feature of D-CryptO, only feature extraction as described earlier was performed. All budding models were trained for 20 epochs and model performance was monitored using the precision metric. All training was done using GPU accessed through Google Colab. Data augmentation was used in all training pipelines to improve model performance by increasing the dataset available to train the model and reducing overfitting. The following data augmentation functions were applied: random flip, random rotation, and random zoom.

**Forskolin treatment**

Forskolin was purchased from STEMCELL Technologies (Cat# 72112). A stock solution of 10mM was prepared following manufacturer instructions. The 10mM stock solution was diluted to 10μM in 1×PBS (Cat#: 14190144). Colorectal organoids were cultured in a 24-well plate for a period of 7-10 days. Forskolin was administered for a period of 2 hours. Brightfield montage images at 4× magnification were taken every 15 minutes using the Cytation 5 cell imaging-multi mode reader.

**Chemotherapeutic drug screen and LDH assay**

Drugs were acquired from the NIH National Cancer Institute at a stock concentration of 10mM diluted in DMSO. Drugs were diluted 200× in Intesticult Medium to achieve a concentration of 50µM. Colon organoids were thawed and embedded in 25µL of Matrigel in a 384 well plate. Following 4 days of culture, drugs were applied. Drug solutions were renewed every other day. Z-stack montages were acquired every other day using a Cytation 5 cell imaging-multi mode reader. The LDH assay was conducted using the CyQUANT LDH Cytotoxicity Assay Kit (Thermo Fisher Scientific, Cat# C20300). First, all required reagents were prepared according to the manufacturer's instructions. Next, media was collected from each drug condition on day 4 and added to a flat-bottom 96-well plate purchased from VWR (Cat# 10062-900). Next, 50µL of the reaction mixture was added to the wells and the plate was incubated for 30 minutes away from light. Following 30 minutes, 50µL of stop solution was added to each well. Absorbance was measured at 490nm and 680nm using the Cytation 5 cell imaging multi-mode reader. The background absorbance was then subtracted from the 490nm absorbance value.

**Histology staining**

Using 10% formalin, the tissues were fixed for 48 hours. The tissues were then extracted from the wells with a tweezer and then placed in histology cassettes and immersed in 70% ethanol. The tissues were then processed by the MIRC histology Core Facility and stained for Villin (Abcam, Cat#130751), E-Cadherin (Abcam, Cat# ab1416), and Ki67 (Abcam, Cat#16667).

**Quantification analysis**

Individual organoids within image montages were detected using OrgaQuant[17] or had boxes drawn around them manually using the SuperAnnotate software. To assess changes in budding and opacity, all predictions by the final trained models were outputted to a CSV file. For opacity, any organoids which had a classification score of greater than 50% were classified as opaque. For budding, any organoids with a classification score of greater than 50% were classified as budding. The change in budding and opacity was assessed in at least 3 independent samples. To measure diameter, the x coordinates from the bounding boxes of the detected organoids were used. The diameter was obtained from at least 3 independent samples. Confusion matrices and the organoid distribution dot plot were plotted using the Matplotlib library.

**Statistical analysis**

All results are plotted as mean ± standard deviation. Normality and equal variance were tested using GraphPad. A p-value < 0.05 was considered statistically significant in all experiments. At least three independent samples were used for all experiments. For data in **Figures 3h-j**, statistical significance was determined using a one-way repeated measures ANOVA followed by Dunnett's test. Statistical significance for **Figures 3c-e** was assessed using an unpaired two-tailed t-test. For Figure 4, statistical significance was assessed using a one-way repeated measures ANOVA followed by Dunnett's test. Statistical significance for changes in diameter following treatment with Erlotinib (Figure 4g) and the changes in opacity in the organoids treated with doxorubicin (Figure 4h) were assessed using the Friedman test followed by Dunn's test. Statistical significance between day 10 opacity, budding, and diameter values to day 10 control values were determined using a two-tailed unpaired t-test. Statistical significance in the opacity of the organoids to the day 10 control in the organoids treated with doxorubicin was assessed using the Mann-Whitney test.

**Data availability**

All the trained models and the datasets can be downloaded from: https://osf.io/42r3g/

**Supporting Information**

Supporting Information is available from the author.

**Acknowledgements**

This work was funded by the National Sciences and Engineering Research Council of Canada (NSERC) Discovery Grant (RGPIN-05500-2018), and the Canadian Institute of Health Research (CIHR) Project Grant (PJT-166052) to BZ. This work was made possible by the financial support of the Canada Graduate Scholarship to L.A. The authors are grateful to BioRender.com which we have used to make the illustrations in this work.

**Author contribution**

L.A. performed the experiments, deep learning analysis, and prepared the manuscript. J.X. helped with dataset creation and model training. A.S. contributed to the culturing of colon organoids. A.C. and J.G. contributed to the dataset creation. S.R. contributed to colon organoid culture. N.A. and H.M. reviewed and edited the manuscript. B.Z. envisioned the concept, supervised the work and edited the manuscript.

**Competing financial interests**

None.

**Table of contents**

This work focuses on the development and validation of D-CryptO, a deep learning-based image analysis tool that can be used to analyze colon organoid structural maturity directly from brightfield images. D-CryptO can detect changes in organoid morphology over prolonged culture, during short-term perturbation, and following chemotherapeutic drug treatment.

Lyan Abdul[1], Jocelyn Xu[2], Alexander Sotra[1], Abbas Chaudary[3], Jerry Gao[4], Shravanthi Rajasekar[3], Nicky Anvari[1], Hamidreza Mahyar[5], and Boyang Zhang[1,3*]

**Title: D-CryptO: Deep learning-based analysis of colon organoid morphology from brightfield images**

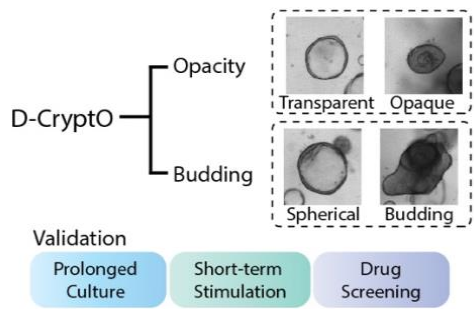

Supporting Information

**D-CryptO: Deep learning-based analysis of colon organoid morphology from brightfield images**

Lyan Abdul[1], Jocelyn Xu[2], Alexander Sotra[1], Abbas Chaudary[3], Jerry Gao[4], Shravanthi Rajasekar[3], Nicky Anvari[1], Hamidreza Mahyar[5], and Boyang Zhang[1,3*]

**Dose-dependent changes in organoid morphology**

To further validate D-CryptO, we used it to assess the dose-dependent response of doxorubicin on opacity and budding following treatment at various concentrations. Doxorubicin is a chemotherapeutic that inhibits DNA and RNA synthesis and induces apoptosis.[34] We applied doxorubicin at concentrations of 50 µM, 5 µM, 0.5µM, 0.05 µM, and 0.005 µM **(Figure S1a)**. For opacity, the concentration at which 50% of organoids became opaque was 3.6 µM **(Figure S1b)**. For budding, the concentration at which 50% of the organoids still had budding structures was 39.8 µM **(Figure S1c)**. It is important to note that budding did not increase with higher dosages of doxorubicin. Instead, the percentage of non-viable organoids increased which was classified under the budding category. For diameter, the concentration at which 50% of the organoids had a reduction in diameter was 0.5 µM **(Figure S1d)**. Each parameter was impacted at different concentrations, indicating the importance of monitoring these features to assess drug toxicity.

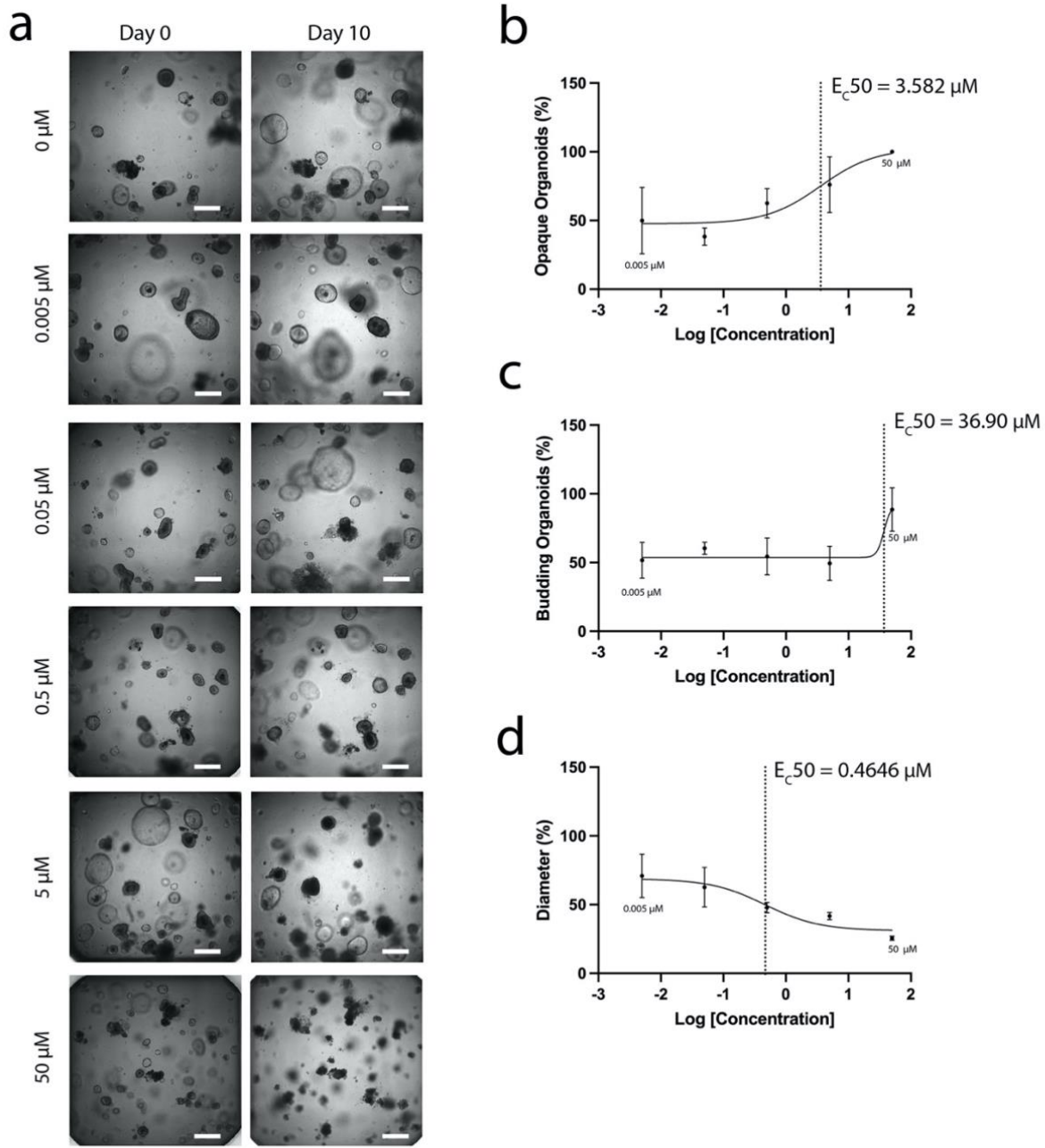

**Supplementary Figure 1. Dose-dependent changes in organoid morphology. a,** Brightfield images of organoids taken on Day 0 and Day 10 of drug treatment with doxorubicin at 5 different concentrations. Scale bar, 500 μm **b,** The percentage of opaque organoids following 10 days of treatment with increasing concentrations of doxorubicin. **c,** The percentage of budding organoids following 10 days of treatment with increasing concentrations of doxorubicin. **d,** The change in diameter following 10 days of treatment with increasing concentrations of doxorubicin.

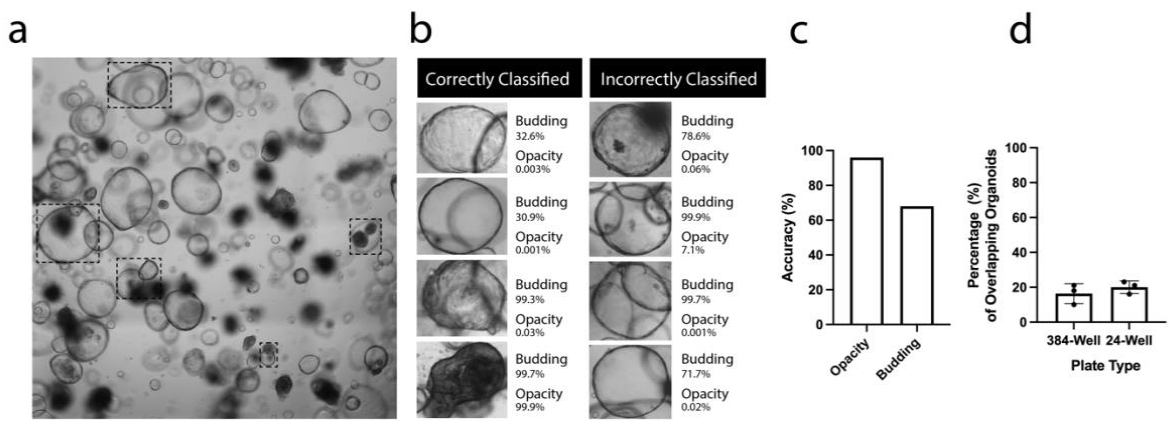

**Supplementary Figure 2. D-CryptO performance on overlapping organoids. a,** Examples of overlapping organoids. **b,** Correct and incorrect classifications by D-CryptO. **c,** D-CryptO accuracy when classifying overlapping organoids. **d,** Percentage of overlapping organoids in both 384 and 24-well plates.

29